**Title: Resonant Silicon Nanoparticles for Enhanced Light Harvesting in Halide Perovskite Solar Cells**

*A.D. Furasova\*[1], E. Calabró[2], E. Lamanna[2], E.Y. Tiguntseva[1], E. Ushakova[1], E. V. Ubyivovk[3], V. Y. Mikhailovskii[3], A.A. Zakhidov[1,4], S.V. Makarov[1], A. Di Carlo\*[2]*

[1]A.D. Furasova, E.Y. Tiguntseva, Dr. E. Ushakova, Prof. A.A. Zakhidov, Dr. S.V. Makarov, ITMO University, St. Petersburg, 197101, Russia
[2]E. Calabro, E. Lamanna, Prof. A. Di Carlo
C.H.O.S.E. (Centre for Hybrid and Organic Solar Energy), Department of Electronic Engineering, University of Rome-Tor Vergata, Rome, 00133, Italy
[3]E. V. Ubyivovk, V. Y. Mikhailovskii
Saint-Petersburg State University, Saint-Petersburg, 199034, Russia
[4] A.A. Zakhidov
University of Texas at Dallas, Richardson, Texas 75080, USA

E-mail: aleksandra.furasova@metalab.ifmo.ru; aldo.dicarlo@uniroma2.it

**Abstract** Implementation of resonant colloidal nanoparticles for improving performance of organometal halide perovskites solar cells is highly prospective approach, because it is compatible with the solution processing techniques used for any organic materials. Previously, resonant metallic nanoparticles have been incorporated into perovskite solar cells for better light absorption and charge separation. However, high inherent optical losses and high reactivity of noble metals with halides in perovskites are main limiting factors for this approach. In turn, low-loss and chemically inert resonant silicon nanoparticles allow for light trapping and enhancement at nanoscale, being suitable for thin film photovoltaics. Here photocurrent and fill-factor enhancements in meso-superstructured organometal halide perovskite solar cells, incorporating resonant silicon nanoparticles between mesoporous $TiO_2$ transport and active layers, are demonstrated. This results in a boost of the device efficiency up to 18.8% and fill factor up to 79%, being a record among the previously reported values on nanoparticles incorporation into $CH_3NH_3PbI_3$ ($MAPbI_3$) perovskites based solar cells. Theoretical modeling and optical characterization reveal the significant role of Si nanoparticles for increased light absorption in the active layer rather than for better charge separation. The proposed strategy is universal and can be applied in perovskite solar cells with various compositions, as well as in other optoelectronic devices.

The organic-inorganic perovskite-based solar cells (PSCs) are the most notable scientific milestone in photovoltaic science and technology of the last years[1-3]. From first works started in 2009[4] to recent developments, solution-processed PSCs have rapidly achieved remarkable power conversion efficiency (PCE) values exciding 22% through a careful materials design and fabrication process optimization[5]. Concerning optimization strategies, the thickness of the perovskite layer plays a key role in light absorption and in charge injection into the hole and electron transport layers[6]: a thin perovskite layer has poor light harvesting but good charge separation and collection capabilities, while the series resistance grows in with thickness of the perovskite film. For this reason, the increase of light harvesting properties of the absorbing perovskite layer without increasing its thickness is an effective strategy to improve PSC performance. In this context, a large attention was paid to the use of the effects related to surface plasmons (resonant light-induced collective excitations of the electron gas[7]) in nanoscale metallic objects incorporated in PSC, providing strong localization of the electromagnetic (EM) field on deeply subwavelength scale for small particles as well as multifold increase of scattering cross-section for larger particles. As a result, the excited surface plasmons increase the absorption in a photo-active layer and, thus, enhance light harvesting in a solar cell[8].

Incorporation of noble metal nanostructures have been considered, beside PSCs, in many PV technologies such as dye sensitized solar cells[9-11], silicon solar cells[12,13], and organic photovoltaic cells[14-16]. Concerning PSCs, the first study on incorporation of metal nanoparticles (NPs) into the perovskite device to exploit the plasmonic effects was carried out by Zhang et al.[17] attracting significant attention to this approach. **Table 1** shows the most remarkable results on incorporation of plasmonic NPs into PSCs based on $CH_3NH_3PbI_3$ ($MAPbI_3$) perovskite with different device architectures.

**Table 1.** Characteristics of $MAPbI_3$ based solar cells with incorporated resonant nanoparticles

| Resonant nanoparticles, shape (diameter) | Device structure | Nanoparticles location | Efficiency (%)/ Fill factor (%) | Ref. |
|---|---|---|---|---|
| Si, nanospheres (100-200 nm) | regular, mesoporous electron transport layer (m-ETL) | between m-ETL and perovskite | 18.8/79 | current work |
| Au@$TiO_2$, nanospheres (80nm) | regular, m-ETL | in m-ETL or/and in perovskite | 18.2/75.5 | [21] |
| Au@$SiO_2$, nanospheres (14 nm) | regular, m-ETL | between compact and m-ETLs | 17.6/78.2 | [19] |
| Au@$SiO_2$, nanorods (15×37 nm) | Inverted | between hole transport layer (HTL) and perovskite | 17.6/77.3 | [24] |
| Au@$TiO_2$, nanorods (5×40) nm | regular, planar | in ETL | 16.8/74.7 | [25] |
| Au, nanospheres (40 nm) | regular, m-ETL | in ETL | 16.2/76 | [22] |
| Au, nanospheres (40 nm) | regular, with MgO layer | between m-ETL and MgO | 16.1/68 | [26] |
| Au@$TiO_2$, nanospheres (60 nm) | regular, with $TiO_2$ nanofibers | on $TiO_2$ nanofibers | 14.9/70 | [18] |
| Au@$SiO_2$, nanospheres (45 nm) / nanorods (8×55 nm) | Inverted | in perovskite | 13.7/68 | [27] |
| Ag@$TiO_2$, nanospheres (40 nm) | regular, m-ETL | between m-ETL and perovskite | 13.7/67 | [20] |
| Au, nanostars (30 nm) | inverted/regular, planar | in HTL | 13.7/72.1 (regular) 8.7/71.2 (invert) | [28] |
| Au@$SiO_2$, nanospheres (100 nm) | regular, m-ETL | in m-ETL | 11.4/64 | [17] |

According to the works cited in **Table 1**, experimental results demonstrated that gold and silver NPs[17-20] in MAPbI$_3$-PSCs influence the photocurrent and photovoltage[19], resulting in improving the final PCE up to 18.2% with a fill factor of 75.5% for the best device[21]. The incorporation of NPs into transport layers or deposition on the interfaces with a perovskite layer are the most popular approaches, because, in these cases, the PCS geometry and perovskite layer morphology are not changed significantly. Moreover, Yuan and co-workers[22] noted that direct plasmonic NP's incorporation into compact electron transport layer (ETL) enhances efficiency by hot-electron injection into ETL. However, high optical losses in metal NPs[23] and the high chemical reactivity of noble metals with halides in perovskites[17, 22] are main limiting factors for the plasmonic approach.

In contrast to metallic NPs, all-dielectric nanophotonics based on silicon (Si) NPs with Mie resonances in visible and infrared ranges recently has emerged as the powerful tool for various optical applications[29, 30]. Owing to their low cost, chemical and temperature stability, Si NPs represent a viable and better alternative to noble metals for photovoltaics application.

Here, for the first time to our knowledge, resonant Si NPs are employed to improve halide PSCs performance, boosting the final PCE from 17.7% (without NPs) up to 18.8% (with NPs), higher than previously reported MAPbI$_3$ based PSCs with plasmonic NPs. This approach opens the way for novel optimization strategies to improve perovskite photovoltaics, by exploiting Mie-enhanced absorption in a photo-active layer by means of stable, chemically inert, low-cost and sustainable Si NPs as compared to noble metal NPs.

Meso-superstructured PSCs[31] are fabricated considering the n-i-p architecture. A 50-nm thick TiO$_2$ compact layer is deposited on top of an electron collecting fluorine-doped tin oxide (FTO) electrode, followed by a mesoporous electron-transport TiO$_2$ 200-nm thick layer, capped by MAPbI₃ perovskite, which forms the photoactive layer, where electrons and holes are generated and separated under solar irradiation. The hole transport SPIRO-OMeTAD layer doped with Bis(trifluoromethane)sulfonimide lithium salt (Li-TFSI), coating perovskite and 80-nm gold layer, was used as HTL and top contact, respectively. Details of the devices creation process are given in the **Experimental Section**. The energy-level diagrams of the component materials are illustrated in **Figure 1 a**. **Figure 1 b** presents a cross-section scanning transmission electron microscopy (STEM) image of the device corresponding to a typical meso-superstructured PSC.

Colloidal Si NPs are deposited on the top of m-TiO₂ by spin coating. The Si NPs fabricated by laser ablation[32] of a silicon wafer in toluene (see the **Experimental Section**) have average size around 140 nm with relatively broad size distribution from 90 to 200 nm at $1/e^2$-level (**Figure S1 a**). According to STEM (**Figure 1 c**) measurements, Si NPs are polycrystalline with almost spherical shape. **Figure S1 b** presents the Si NPs distribution on a m-TiO₂ layer after their deposition. According to **Figures S2 and S3**, Si NPs located between m-TiO₂ and perovskite layers do not affect the layers morphology, making the perovskite film as smooth as that without NPs.

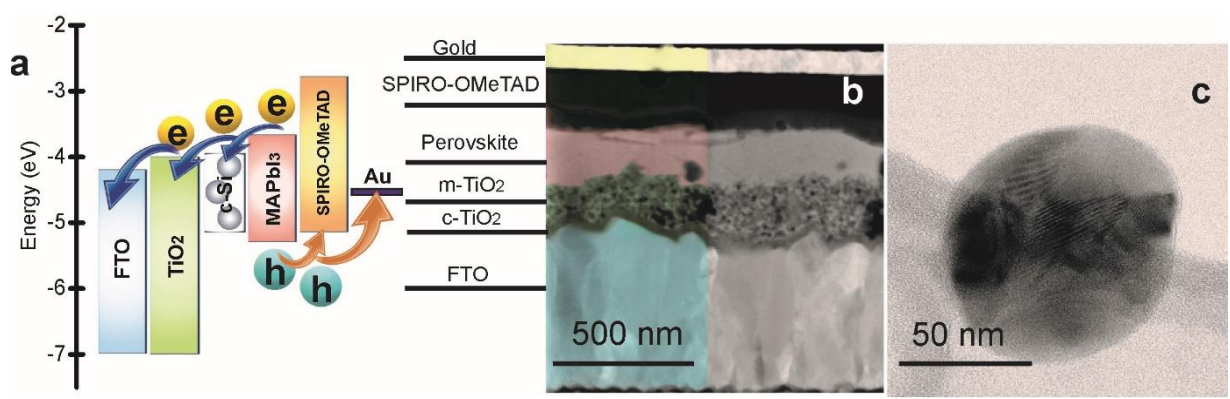

**Figure 1.** (a) Energy band diagram of the device. (b) Color-enhanced and annotated cross-sectional bright-field STEM image of a complete solar cell. (c) TEM image of a polycrystalline Si nanoparticle fabricated by laser ablation in liquid.

According to Mie theory[33, 34], Si NPs with diameters more than 100 nm can support magnetic and electric type resonances in visible and infrared ranges, yielding high scattering efficiencies values in air[35], liquids[36], or perovskites[37]. Indeed, the real part of the refractive index of polycrystalline silicon (n ≈ 3–4) is significantly higher than that of MAPbI$_3$ (n ≈ 2.5) and titanium dioxide (TiO$_2$ with n ≈ 2.5) in the visible range. Therefore, the optical contrast is enough to excite the optical resonances in the Si nanosphere even embedded between MAPbI$_3$ and TiO$_2$ layers. According to our analytical calculations based on the Mie theory and described in SI, the chosen diameters of Si NPs (in the rage of 100-200 nm) allow for excitation of Mie type optical resonances (**Figure 2 a**), namely, magnetic and electric dipole resonances, as well as electric and magnetic quadrupole resonances which are partially overlapped spectrally. Such a broadband resonant behavior in visible range can provide a near-field enhancement inside the perovskite film and increased scattering inside the film, resulting in better light absorption.

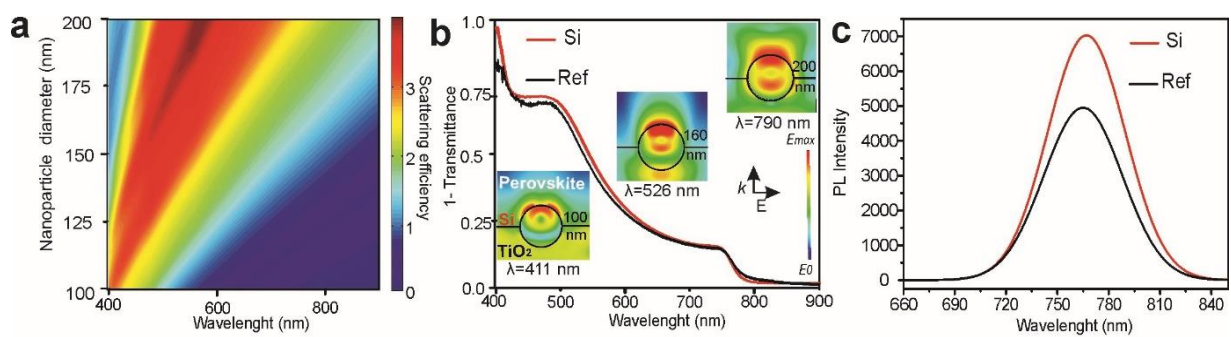

**Figure 2.** (a) Mie theory calculation of scattering efficiency as a function of diameter of a silicon nanoparticle in bulk $TiO_2$ host medium and wavelength of incident light. (b) Experimental extinction (1-T) spectrum of the perovskite solar cells without a top contact: for solar cell without (black line) and with (red line) Si NPs. Insets: numerical calculations of electric near-field enhancement for the devices with silicon NPs of different sizes and different wavelengths, pointed respectively. (c) Photoluminescence (PL) intensity from a perovskite layer of the device (without SPIRO-OMeTAD and top contact): for a reference (black line) and with Si (red line) NPs.

Optical properties of Si NPs located between m-$TiO_2$ and $MAPbI_3$ layers, were studied by numerical simulations using the CST Microwave Studio software (for details, see Experimental Section). The calculations allow us to estimate the electric field enhancement around the interfaces with and without NPs, and to find the relation between NPs size and the maximum enhancement of optical field (**Figure 2 b**). The calculated electric field distribution shows that Si NPs with diameter in the range of 140-160 nm exhibit a scattering efficiency maximum in the green range, which corresponds to highest intensity of the sun light spectrum, whereas 200-nm Si NPs contribute more in the red range, where absorption of $MAPbI_3$ is lower. Additional numerical modeling (see **Figure S4**) reveals near-field enhancement around ensemble of randomly distributed Si NPs of different sizes (taken from our dynamic light scattering measurements) at the interface between $TiO_2$ and perovskite layers.

Experimental part of Figure 2 b shows the sum of absorbance, scattering and reflectance presented as *1-Trasmittance*, for FTO/c-$TiO_2$/m-$TiO_2$/Perovskite samples with (red curve) and without (black curve) Si NPs with the size distribution shown in Figure S1.

The device with Si NPs is less transparent in the visible range than the reference one. This is in qualitative agreement with the calculations showing a more effective near-field enhancement around the nanoparticles, i.e. light trapping. According to calculations of absorption cross-section of Si nanoparticle in $TiO_2$ material (**Figure S5**), the spectral range 400-500 nm corresponds to enhanced absorption in the nanoparticle, owing to high absorption coefficient of silicon at the photon energies close to direct interband transition (3.4 eV). Thereofore, enhanced light absorption between 400 and 500 nm by the perovskite layer in our devices (Figure 2 b) can be partially attributed to the absorption in Si NPs. For wavelengths larger than 500 nm, the absorption in Si NPs can not contribute significantly to the total absorption and, thus, Si NPs work effectively as nanoantennas in the range $\lambda>500$ nm. We also studied the photoluminescence (PL) at room temperature (**Figure 2 c**) and observed the increase of the PL signal for the devices with Si NPs. PL enhancement is homogeneous over the film area, as shown in **Figure S6**. Moreover, we registered even longer decay times for our devices with Si NPs, as shown in **Figure S7**.

These experimental results are opposite to those reported for incorporation of metallic NPs into PSCs, where PL signal was observed to be quenched, whereas absorption of incident light does not change significantly [17,21]. Such PL quenching by plasmonic NPs usually results in acceleration of PL decay time, which was observed previously for various light-emitters placed nearby metallic objects[38, 39]. This process is usually related to the increase of PSCs efficiency, owing to enhanced charge separation[18, 22]. In turn, as described in previous theoretical papers[40, 41], the role of resonant plasmonic NPs is to increase the light absorption without contributing to the charge separation. In particular, numerical simulations for plasmonic NPs showed that randomly located plasmonic NPs in a photoactive perovskite ($MAPbI_3$) layer can enhance the light absorbance up to 1.12 times, revealing the optimal particle sizes[40]. However, the presence of low-loss dielectric NPs in PSCs should not induce the PL quenching, and PL intensity tends to be enhanced due to the local optical field enhancement in the perovskite layer. Indeed, this is the result of fundamental principle of the detailed balance, where the external luminescence exactly balances the incoming sunlight, if the nonradiative loss channels are not changed[42], which is the case of our low-loss Si NPs resonantly enhancing both absorption of incoming sunlight and PL from the thin perovskite layer (Figures 2 b,c).

A set of 30 similar devices was separated to two independent batches of production to assess the influence of Si NPs on the photovoltaic parameters. As shown in **Figure 3**, all PSCs (with and without Si NPs) show a PCE higher than 15%, which corresponds to the standard state-of-the-art fabrication process for $MAPbI_3$ based PSCs. The champion reference device showed a PCE of 17.7% that is one of the highest results for $MAPbI_3$ based PSC[43]. As shown in Figure 3 d, the presence of Si NPs clearly increases the average and the best efficiency of the cells with respect to that without NPs. In **Figure 4 a**, the current density–voltage (J-V) scans for the best reference PSC together with the best PSC with Si NPs are shown. The corresponding PV parameters including short-circuit current density ($J_{SC}$), open circuit voltage ($V_{OC}$), fill factor (FF) and PCE are summarized in **Table 2**. A clear increase of PCE from 17.7% (without Si NPs) to 18.8% (with Si. NPs) is observed and it is related to an increase of $J_{SC}$ (+1.8%), FF (+2.4%), and $V_{OC}$ (+1%).

Generally, the open-circuit voltage is the voltage at which the forward bias diffusion current is equal to the short circuit current. The forward bias diffusion current is dependent on the amount recombination and increasing the recombination increases the forward bias current. Consequently, high recombination increases the forward bias diffusion current, which in turn reduces the open-circuit voltage. As a result, $V_{OC}$ as well as PL and time-resolved PL (see Figure S7) are dependent on defects concentration[42,44]. This idea is supported by the observation of some improvements of perovskite crystallinity in our XRD measurements (for details, see **Figure S8** and **Table S2**). However, 50% enhancement of PL can not be attributed to the improved crystallinity only, and, thus, Mie resonances in Si NPs at the emission wavelength (770 nm) provide additional contribution via Purcell effect [45]. The short-circuit current density is increased due to the local field enhancement around Si NPs in broad spectral range, resulting in enhanced optical absorption and generation of larger number of electron-hole pairs.

**Figure 4 b** shows the comparison of external quantum efficiency (EQE) between devices with and without Si NPs. EQE is clearly increased in the range between 500 and 800 nm in the devices with Si NPs with respect to those without Si NPs. Moreover, EQE exhibits the strongest enhancement in the red region (550-770 nm, inset of Fig. 3 b), owing to the lower inherent losses of Si NPs and perovskite, yielding stronger near-field enhancement (Figure S4). We should point out that such efficiency enhancement is achieved together with PL enhancements (shown in Figures 2 c and S6) supporting the mechanism of light trapping around Si NPs predicted by our modeling (Figure 2 b), rather than better carriers separation or by charge injection as it is usually described for plasmonic NPs[20,21]. Importantly, that the light trapping effect almost vanishes for the case, when Si NPs are inside perovskite layer (**Figure S9**), i.e. $J_{SC}$ is similar to that for the devices without the nanoparticles, whereas $V_{OC}$ and FF are similar to those for the devices with Si NPs. In this case, light absorption in the perovskite reduces amount of photons interacting with Si NPs in bulk material.

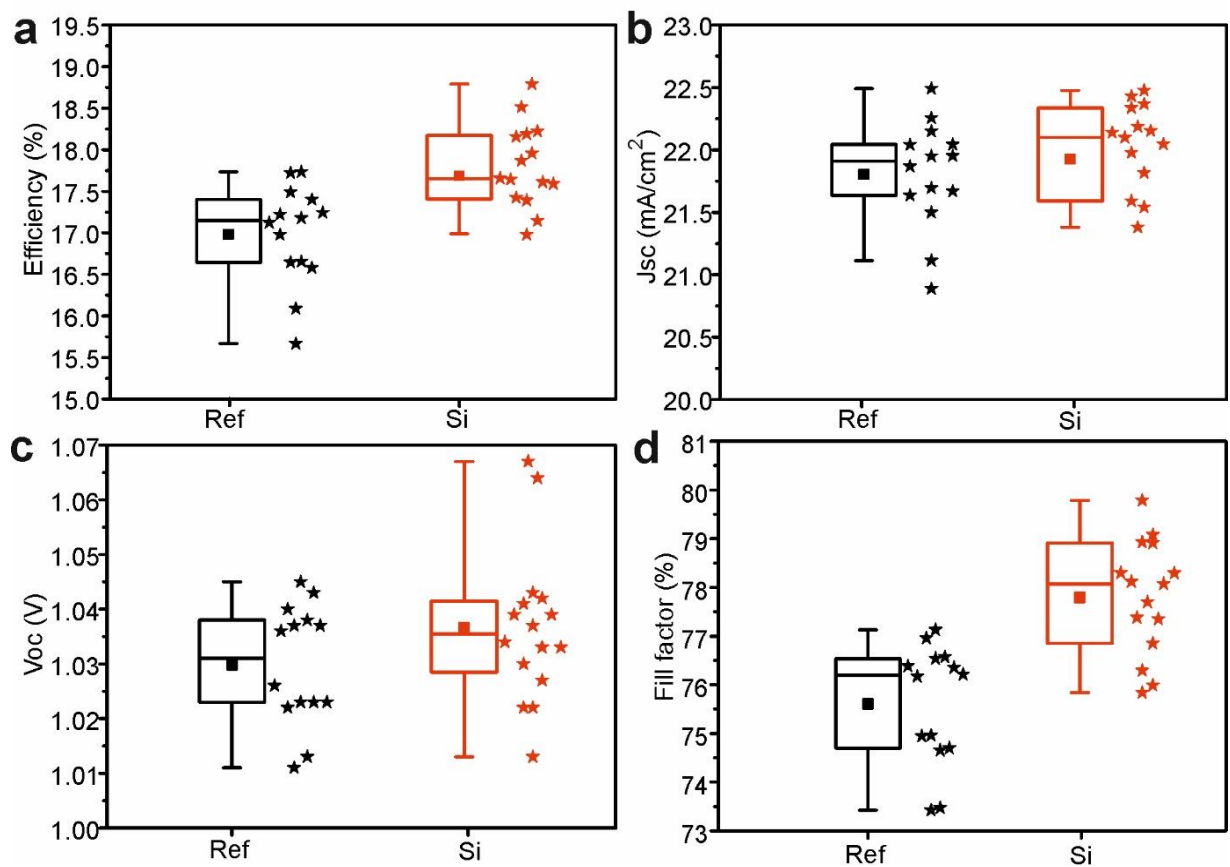

**Figure 3.** Statistical data for photovoltaic performance of individual devices under 1 sun illumination (black: reference cells, red: cells with Si NPs). The error bars represent a standard deviation from the mean values. (a) Efficiency, (b) short circuit current density, (c) open circuit voltage and (d) fill factor plotted for cell with and without Si NP's interlayer.

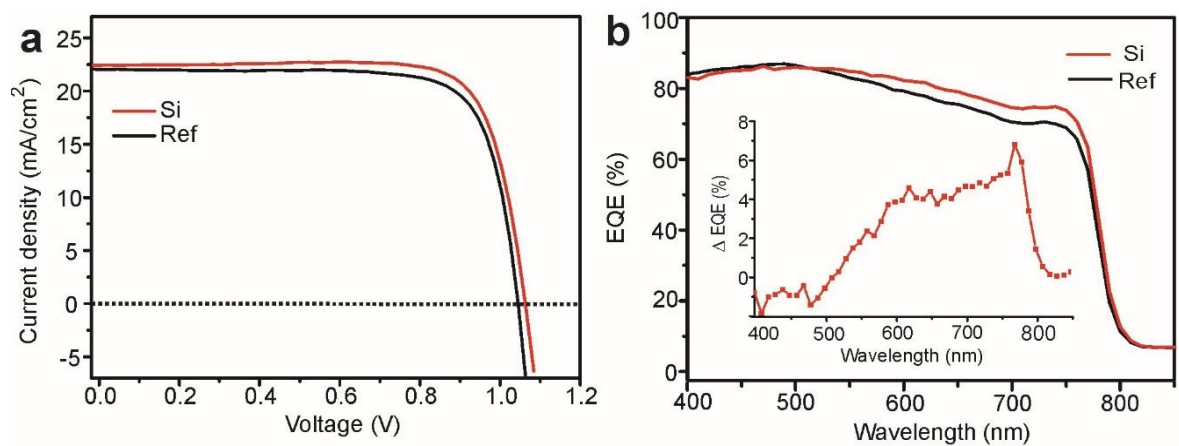

**Figure 4.** (a) J–V curves of the champion solar cells: reference device (black line) and devices with Si NPs (red line) measured at AM1.5G and 100 mW·cm$^{-2}$ at room temperature in air condition. (b) Incident external quantum efficiency (EQE) curves for devices without (black line) and with (red line) Si NPs. Inset: the difference between external quantum efficiencies of the devices without (black line) and (red line) Si NPs in PSC.

**Table 2.** Values of $V_{OC}$, $J_{SC}$, FF, and efficiency of the best perovskite solar cells with and without Si NPs under 1 sun illumination intensity.

| Samples | $J_{SC}$(mA/cm$^2$) for the champion/ average | $V_{OC}$(V) for the champion/ average | FF(%) for the champion/ average | Eff (%) for the champion / average |
|---|---|---|---|---|
| Reference (champion) | 22.0/ 21.8 | 1.05/1.03 | 77.0/75.6 | 17.7/17.2 |
| Si (champion) | 22.4/ 22.2 | 1.06/1.04 | 78.9/77.6 | 18.8/17.7 |

The PCE results discussed above are related to the reverse scan of the J-V curves. However, for the mesoscopic PSCs the hysteresis effect cannot be neglected. It is well known that the hysteresis index (HI) is quantitative parameter to determine the difference in power output depending on scan direction[46]. In our case, calculations of HI by using Eq. S5 with the J-V forward and reverse scans of **Figure S10** give HI=0.0954 for the best reference and HI=0.0748 for the best device with Si NPs. The HI decrease after the Si NPs incorporation might be related to the improving of a crystalline structure[47] of MAPbI$_3$. As an additional positive effect of Si NPs on PSC, we have observed that the their presence improves the device performance after 4 days on the shelf and, thus, stability of the PSCs as shown in **Figure S11** and discussed in Supporting Information.

In summary, we have proposed a novel approach to improve the performance of lead halide perovskite solar cells by incorporation of resonant silicon nanoparticles at an interface of MAPbI$_3$ and m-TiO$_2$. The improved devices have demonstrated high short circuit current (up to $J_{SC}$ = 22.4 mA/cm$^2$), fill factor (up to FF = 78.9%), and efficiency (up to Eff = 18.8%), which are higher than those for the reference devices without the nanoparticles and all previously reported plasmonic approaches. While the increase of $J_{SC}$ is originated from increased light absorption around Si NPs, the improved fill factor and the open-circuit voltage can be related to improved crystallinity of the perovskite layer in presence of Si NPs. Taking into account the low cost, high thermal and chemical stability of colloidal silicon nanoparticles, our approach paves the way toward efficient and cheap method for improvement of perovskite-based solar cells. We believe that resonant silicon nanoparticles can be applied not only in solar cells but also in other optoelectronic devices such as LEDs and lasers, owing to overall improvement of light localization and optical field enhancement by Si NPs[48]. Also, as an outlook of our work, we believe that resonant Si NPs will be incorporated further into SCs based on halide perovskites with mixed anions and cations[49] to boost further their efficiencies[25].

**Experimental Section**

*NPs synthesis*:

Si nanoparticles were fabricated by laser ablation method in liquid according to the procedure mentioned in Ref.[44]. A commercial femtosecond laser system (Femtosecond Oscillator TiF-100F, AvestaPoject) was used, generating laser pulses at 800 nm central wavelength, with maximum pulse energies up to 5 mJ, and pulse duration 40 fs at repetition rate 1 kHz. The laser energy was varied and controlled by an acousto-optical modulator (R23080-3-LTD, Gooch and Housego) and a power meter (FielfMax II, Coherent), respectively, while the pulse duration was measured by an autocorrelator (AvestaPoject). Laser pulses were focused by a lens with a focal length of 5 cm on a 500-μm thick Si wafer covered by approximately 2 mm toluene layer. For Si sample, the near-threshold regime of ablation was chosen to provide larger particles formation. Concentration of Si NPs colloid was measured approximately $3 \cdot 10^{-9}$ mol/L.

*Device preparation*:

All procedures, excluding SPIRO-OMeTAD deposition, were held in an atmosphere condition. The FTO substrates were cleaned from heavy pollution by soap and DI water, then, were soaked 3 times for 10 minutes in water with a soap, acetone and isopropanol. Before c-$TiO_2$ deposition, Ag paste coated the bottom contact area, and the substrates were heated from room temperature to 450 ºC. A compact layer of $TiO_2$ was deposited 15 times with the 10 sec interval between each deposition on the heated to 450°C FTO substrates using spray pyrolysis. The spray solution was composed of ethanol, acetyl acetone, and titanium diisopropoxide (30% in isopropanol) in the proportions 90:4:6 by volume. Atmosphere gas at a base pressure of 1.5 bar was used as a carrier gas. After that, the sprayed glass samples were slowly cooled to room temperature. The complete compact $TiO_2$ layer of anatase has a thickness of 50 nm. The mesoporous $TiO_2$ layer was deposited by spin-coating on top of the c-$TiO_2$. For this purpose a $TiO_2$ paste (Dyesol) was dissolved in ethanol as 1 to 5 in a mass. The $TiO_2$ solution was dropped and spin-coated at 3000 rpm for 20 s. After that, the contact area was carefully cleaned from $TiO_2$ layer by acetone. The substrates with FTO, c- and m-$TiO_2$ were calcined at a following temperature regime: 120ºC (10 min) - 325ºC (5 min) - 375ºC (5 min) - 500ºC (30 min) on a hot plate and then cooled to room temperature. To deposit an interlayer of Si NPs, toluene solutions was used. The NPs solutions were deposited by spin-coating method on a m-$TiO_2$ layer at 1000 rpm for 10 s and then heated at 70ºC on a hot plate in the air. In advance prepared $MAPbI_3$ solution (1.4 mmol/ml) in DMF:DMSO (9:1) was spin-coated in a two-step program at 1000 and 5000 rpm for 10 and 45 s, respectively. During the second step, 0.7 ml of diethyl ether was poured on the spinning substrate. The perovskite films after spin-coating immediately turned to hot plate for annealing at 50°C for 1.5 min and then at 100°C for 10 min in darkness at the air atmosphere. Then, devices with completed perovskite layer were put to glove box and SPIRO-OMeTAD doped Bis(trifluoromethane)sulfonimide lithium salt (Li-TFSI) solution (60 mM in chlorobenzene) was spin-coated at 2000 rpm for 15 s in the $N_2$ atmosphere. Finally, the top contacts were composed of an 80 nm thick gold film deposited by physical vapor deposition at a pressure of around $9·10^{-6}$ Torr with a 0.3 Å/s rate for first 10 nm and then 1 Å/s rate up to 80 nm thickness using an evaporator from Kurt Lesker.

*Characterization and Measurements:*

Cross-section of samples for TEM and STEM were prepared by $Ga^+$ focused ion beam (FIB) Zeiss Auriga Laser. For rough milling accelerating voltage 30 kV was used, the ion probe current was 4 nA. Fine milling was performed by low-current ion probe 1 pA. Low accelerating voltage 2 kV was used at final polishing step. The thickness of TEM lamella was about 60 nm. Samples were investigated with TEM Zeiss Libra 200FE at accelerating voltage 200 kV. The absorption spectra were performed by UV3600 spectrophotometer (Shimadzu) in a region from 300 to 1500 nm. The J-V measurements of the PSCs were carried out in air environment under a solar simulator (ABET Sun 2000) at AM1.5G and 100 mW·cm$^{-2}$ illumination conditions, calibrated with a certified reference Si Cell (RERA Solutions RR-1002). Incident light power was measured with a Skye SKS 1110 sensor. The active area was defined through an aperture mask with an area of 0.09 cm$^2$. The scan rate was 30 mV s$^{-1}$ without any preconditioning. The external quantum efficiency (EQE) was performed by using an commercial apparatus (Arkeo – Cicci Research s.r.l.) with an integrated Xenon lamp, a monochromator (Newport 74000) and a solid national Instrument board to measure the photocurrent. Steady state PL measurements were performed with a homemade apparatus composed by a 0.3 m focal length spectrograph with a photon counting unit. The substrates were excited by a green (532 nm) laser at 45° of incidence with a circular spot diameter of 1 mm. The optical coupling system is composed by a lens condenser and a long pass filter. Cross-section of the devices was studied with a scanning electron microscope (SEM, Carl Zeiss, Neon 40), whereas nanoparticle structure and shape was analyzed with transmission electron microscope (TEM, Zeiss Libra 200FE). PL decay at room temperature has been investigated by a laser scanning confocal microscope MicroTime 100 (PicoQuant) equipped with Plan 3.5x, NA=0.10 (LOMO) and Plan N 20x, NA=0.40 (Olympus) objectives and picosecond pulsed diode laser head ($\lambda$ = 405 nm), which implements the method of time-correlated single photon counting.

**Supporting Information**
Supporting Information is available from the Wiley Online Library or from the author.

**Acknowledgements**
((This work was supported by the Ministry of Education and Science of the Russian Federation (Projects 14.Y26.31.0010, and 16.8939.2017/8.9), and the Welch Foundation grant AT 16-17. We thank the IRC for Nanotechnology of Saint-Petersburg State University for assistance in performing the FIB milling and TEM characterization.))

Received: ((will be filled in by the editorial staff))
Revised: ((will be filled in by the editorial staff))
Published online: ((will be filled in by the editorial staff))

References

[1] M. A. Green, A. Ho-Baillie, H. J. Snaith, *Nat. Photonics*. **2014**, *8*, 7.

[2] M. Grätzel, *Nat. Mater.* **2014**, *13*.

[3] J.-P. Correa-Baena, M. Saliba, T. Buonassisi, M.l Grätzel, A. Abate, W. Tress, A. Hagfeldt, *Science.* **2017**, *358*, 6364.

[4] A. Kojima, K. Teshima, Y. Shirai, T. Miyasaka, *J. Am. Chem. Soc*. **2009**, *131*, 17.

[5] W. S. Yang, B.-W. Park, E. H. Jung, N. J. Jeon, Y. C. Kim, D. U. Lee, S. S. Shin, J. Seo, E. K. Kim, J. H. Noh, S. I. Seok, *Science.* **2017**, *356*, 6345.

[6] D. Liu, M. K. Gangishetty, T. L. Kelly, *J. Mater. Chem. A*. **2014**, *2*, 19873.

[7] S. A. Maier, *Plasmonics: Fundamentals and Applications*, Springer Science & Business Media, **2007**.

[8] H. A. Atwater, A. Polman, *Nat. Mater.* **2010**, *9*.

[9] C. Wen, K. Ishikawa, M. Kishima, K. Yamada, *Sol. Energy Mater. Sol. Cells.* **2000**, *61*, 4.

[10] M. D. Brown, T. Suteewong, R. S. S. Kumar, V. D'Innocenzo, A. Petrozza, M. M. Lee, U. Wiesner, H. J. Snaith, *Nano Lett*. **2011**, *11*, 2.

[11] I-K. Ding, J. Zhu, W. Cai, S.-J. Moon, N. Cai, P. Wang, S. M Zakeeruddin, M. Grätzel, M. L. Brongersma, Y. Cui, *Adv. Energy Mater.* **2010**, *1*, 1.

[12] D. Derkacs, S. H. Lim, P. Matheu, W. Mar, E. T. Yu, *Appl. Phys. Lett*. **2006**, *89*, 093103.

[13] S. Pillaia, K. R. Catchpole, T. Trupke, M. A. Green, *J. Appl. Phys*. **2007**, *101*.

[14] S. Hayashi, K. Kozaru, K. Yamamoto, *Solid State Commun.* **1991**, *79*, 9.

[15] J. H. Lee, J. Hwan Park, J. S. Kim, D. Y. Lee, K. Choa, *Org. Electron.* **2009**, *10*, 3.

[16] I. Vangelidis, A. Theodosi, M. J Beliatis, K. Gandhi, A. Laskarakis, P. Patsalas, S. Logothetidis, S. R. P. P Silva, E. Lidorikis, *ACS Photonics.* **2018**, DOI: 10.1021/acsphotonics.7b01390.

[17] W. Zhang, M. Saliba, S. D. Stranks, Y. Sun, X. Shi, U. Wiesner, H. J. Snaith, *Nano Lett.* **2013**, *13*, 9.

[18] S. S. Mali, C. S. Shim, H. Kim, P. S. Patil, C. K. Hong, *Nanoscale.* **2016**, *8*, 5.

[19] N. Aeineh, E. M. Barea, A. Behjat, N. Sharifi, I. Mora-Sero, *ACS Appl. Mater. Interfaces.* **2017**, *9*, 15.

[20] M. Saliba, W. Zhang, V. M. Burlakov, S. D. Stranks, Y. Sun, J. M. Ball, M. B. Johnston, A. Goriely, U. Wiesner, H. J. Snaith, *Adv. Funct. Mater.* **2015**, *25*, 31.

[21] Q. Luo, C. Zhang, X. Deng, H. Zhu, Z. Li, Z. Wang, X. Chen, S. Huang, *ACS Appl. Mater. Interfaces.* **2017**, *9,* 40.

[22] Z. Yuan, Z. Wu, S. Bai, Z. Xia, W. Xu, T. Song, H. Wu, L. Xu, J. Si, Y. Jin, B. Sun, *Adv. Energy Mater.* **2015**, *5*, 10.

[23] G. V. Naik, V. M. Shalaev, A. Boltasseva, *Adv. Mater.* **2013**, *25*, 24.

[24] R. Wu, B. Yang, C. Zhang, Y. Huang, Y. Cui, P. Liu, C. Zhou, Y. Hao, Y. Gao, J. Yang, *J. Phys. Chem. C.* **2016**, *120*, 13.

[25] R. Fan, L. Wang, Y. Chen, G. Zheng, L. Li, Z. Lia, H. Zhou, *J. Mater. Chem. A.* **2017**, *5*, 24.

[26] C. Zhang, Q. Luo, J. Shi, L. Yue, Z. Wang, X. Chena, S. Huang, *Nanoscale.* **2017**, *9*, 8.

[27] J. Cui, C. Chen, J. Han, K. Cao, W. Zhang, Y. Shen, M. Wang, *Adv.Sci.* **2016**, *3*, 3.

[28] R. T. Ginting, S. Kaur, D.-K. Lim, J. M. Kim, J.-H. Lee, S. H. Lee, J. Kang, *ACS Appl. Mater. Interfaces.* **2017**, *9*, 41.

[29] A. I. Kuznetsov, A. E. Miroshnichenko, M. L. Brongersma, Y. S. Kivshar, B. Luk'yanchuk, *Science.* **2016**, *354*, 6314.

[30] I. Staude, J. Schilling, *Nat. Photonics.* **2017**, *11*.

[31] M. M. Lee, J. Teuscher, T. Miyasaka, T. N. Murakami, H. J. Snaith. *Science,* **2012**, *338*, 6107.

[32] D. Zhang, B. Gökce, S. Barcikowski, *Chem. Rev.* **2017**, *117*, 5.

[33] G. Mie, *Ann. Phys.*, **1908**, *330*, 3.

[34]     C. F. Bohren, D. R. Huffman, *Absorption and scattering of light by small particles*, John Wiley & Sons, Wiley Science Series **2008**.

[35] A. B. Evlyukhin, S. M. Novikov, U. Zywietz, R. L. Eriksen, C. Reinhardt, S. I. Bozhevolnyi, B. N. Chichkov,        *Nano Lett.* **2012**, *12*, 7.

[36] L. Shi, T. U. Tuzer, R. Fenollosa, F. Meseguer, *Adv. Mater.* **2012**, *24*, 44.

[37] E. Tiguntseva, A. Chebykin, A. Ishteev, R. Haroldson, B. Balachandran, E. Ushakova, F. Komissarenko, H. Wang, V. Milichko, A. Tsypkin, D. Zuev, W. Hu, S. Makarov, A. Zakhidov,       *Nanoscale.* **2017**, *9*, 34.

[38] B. N. J. Persson, N. D. Lang, *Phys. Rev. B*. **1982**, *26*, 5409.

[39] E. Dulkeith, A. C. Morteani, T. Niedereichholz, T. A. Klar, J. Feldmann, S. A. Levi, F. C. J. M. van Veggel, D. N. Reinhoudt, M. Möller, D. I. Gittins, *Phys. Rev. Lett*. **2002**, *89*, 203002.

[40]     S. Carretero-Palacios, M. E. Calvo, H. Míguez,       *J. Phys. Chem. C.* **2015**, *119*, 32.

[41]     S. Carretero-Palacios, A. Jiménez-Solano, H. Míguez, *ACS Energy Lett.* **2016**, *1*, 1.

[42] O. D. Miller, E. Yablonovitch, S. R. Kurtz, *IEEE Journal of Photovoltaics*, **2012**, *2*, 3.

[43] G. Yang, H. Tao, P. Qin, W. Ke, G. Fang, *J. Mater. Chem. A*. **2016**, *4*, 3970.

[44] S.D. Stranks, *ACS Energy Lett*. **2017**, *2* 7.

[45] M. V. Zyuzin, D. G. Baranov, A. Escudero, I. Chakraborty, A. Tsypkin, E. V. Ushakova, F. Kraus, W. J. Parak, S. V. Makarov, *Sci. Rep.* **2018**, *8*, 6107.

[46] J. W. Lee, S.G. Kim, S. H. Bae, D. K. Lee, O. Lin, Y. Yang, N. G. Park, *Nano Lett*. **2017**, *17,* 7.

[47] H. S. Kim, N. G. Park, *J. Phys. Chem. Lett.* **2014**, *5*, 17.

[48] S. Makarov, A. Furasova, E. Tiguntseva, A. Hemmetter, A. Berestennikov, A. Pushkarev, A. Zakhidov, Y. Kivshar, Halide-Perovskite Resonant Nanophotonics, arXiv preprint arXiv:1806.08917, **2018**.

[49] D. Bi, W. Tress, M. I. Dar, P. Gao, J. Luo, C. Renevier, K. Schenk, A. Abate, Fabrizio Giordano, J.-P. C. Baena, J.-D. Decoppet, S. M. Zakeeruddin, M. K. Nazeeruddin, M. Grätzel, A. Hagfeldt, *Sci. Adv.* **2016**, *2,* 1.

**The table of contents entry should be 50−60 words long**, and the first phrase should be bold. The entry should be written in the present tense and impersonal style.

**A novel strategy for boosting efficiency of organometal perovskite solar cells** is proposed, basing on resonant silicon nanoparticles incorporated between an photo-active and transport layers. The device efficiency is increased up to 18.8% and fill factor up to 79%, being a record among the previously reported results on nanoparticles incorporation into standard $CH_3NH_3PbI_3$ ($MAPbI_3$) perovskites based solar cells.

**Keywords: organometal halide perovskites, solar cells, silicon nanoparticles, all-dielectric nanophotonics, Mie resonances**

A.D. Furasova*[1], E. Calabro[2], E. Lamanna[2], E.Y. Tiguntseva[1], E. Ushakova[1], E. V. Ubyivovk[3], V. Y. Mikhailovskii[3], A.A. Zakhidov[1,4], S.V. Makarov[1], A. Di Carlo*[2]

**Title**

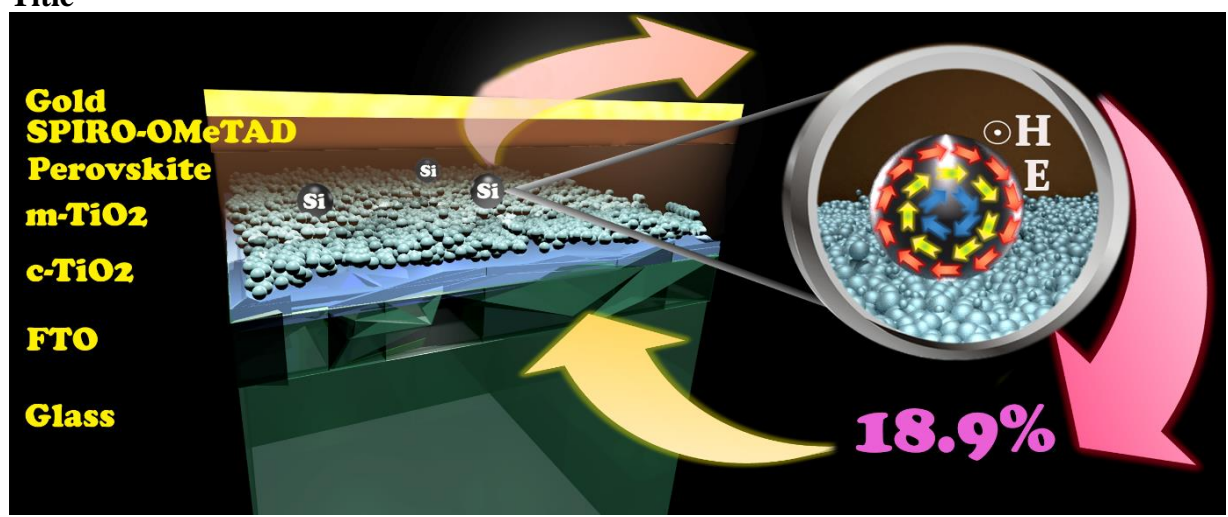